\documentclass[journal=jacsat,manuscript=article]{achemso}

\usepackage[version=3]{mhchem} % Formula subscripts using \ce{}
\usepackage[utf8]{inputenc}
\usepackage[labelfont=bf]{caption}

\author{Marek Foltyn}
\author{Maciej Zgirski}
\email{zgirski@ifpan.edu.pl}
\affiliation[Institute of Physics, Polish Academy of Sciences]
{Institute of Physics, Polish Academy of Sciences,
al. Lotników 32/46, PL 02-668 Warszawa, Poland}

\title[Gambling with superconducting fluctuations]
  {Gambling with superconducting fluctuations}

\abbreviations{}

\begin{document}
\singlespacing
\setlength{\parindent}{1cm}

\begin{abstract}
Superconducting nanowires and Josephson junctions, when biased close to superconducting critical current, can switch to a non-zero voltage state by thermal or quantum fluctuations. The process is understood as an escape of a Brownian particle from a metastable state. Since this effect is fully stochastic, we propose to use it for generating random numbers. We present protocol for obtaining random numbers and test the experimentally harvested data for their fidelity.
\end{abstract}

\subsection{Introduction}
In microworld particles are subject to random interaction with environment and undergo perpetual random walk. Usually these temperature stimulated movements, random and uncorrelated at single particle level, seem to be not visible in everyday live and in fact were first identified only in 1827, by botanist Robert Brown. Subsequently Johnson and Nyquist proved that thermal motions are also responsible for unwanted noise in electrical circuits that competes with a desired signal\cite{Johnson1928,Nyquist1928}. However they are marvelous situations when they manifest themselves in a more sophisticated manner e.g. rubber band holds a stack of paper because smaller molecules “kick” the long ones, thus not allowing them to elongate\cite{bbctv}. Quite recently Brownian motions have been utilized in a modern \textsl{Maxwell’s demon} experiment to transfer “hot” electrons from colder to hotter electrode across a tunnel junction leading to refrigeration of the former\cite{Pekola2011}.

In the current work we propose to use the Brownian behavior of a superconducting wave function of a Josephson junction (JJ) or a superconducting nanowire to generate a sequence of random numbers. In superconductivity electrons are strongly correlated what allows to describe them with a single macroscopic wave function. The phase of the wave function, interacting randomly with environment, fluctuates just like a position of a single particle. This time, however, we deal with a macroscopic Brownian particle, since for any fluctuation to happen, many electrons must be involved. It has been shown that JJs provide a convenient landscape for studies of superconducting fluctuations\cite{Martinis1988, brownian_particle_Esteve}. In the following we briefly describe the roots of randomness in JJs and superconducting wires. We show that our nanostructures exhibit a digital behavior: they may be found in two easily distinguishable, non-arbitrary states, encoding logical 0 and 1. We present a protocol allowing to obtain a random sequence of bits. Finally, we perform a few statistical tests on the experimentally obtained sequences of data, which prove the randomness of our bit series. A short overview of existing random number generators is presented in the Supplemental Material.
\subsection{Josephson junction and superconducting nanowires as random switches}
\begin{figure}
\centering
\includegraphics[width=0.7\textwidth]{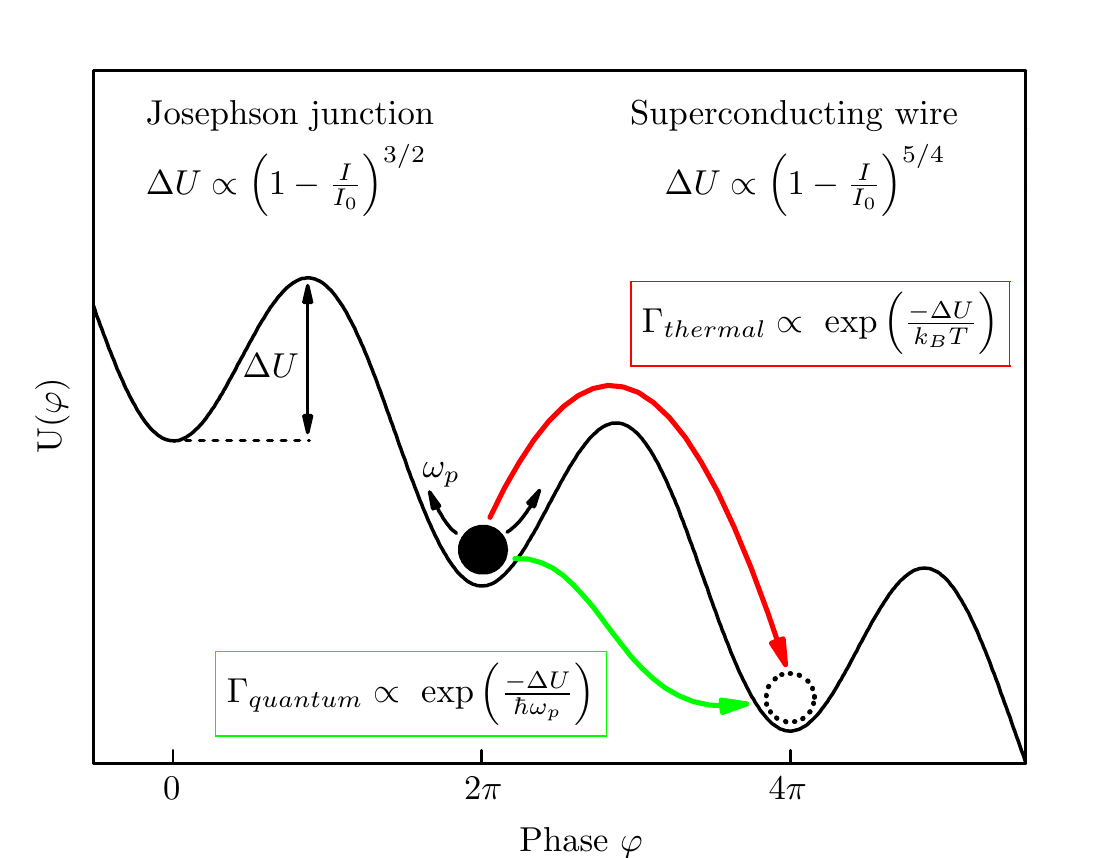}
\caption{Brownian particle undergoing random oscillations in tilted washboard potential can jump over or tunnel through barrier (switching), or may stay trapped in the well (no switching). \(\Gamma\)s denote rates for both processes. The fluctuations of the particle are visualized in Supplemental Material.}
\label{fig:JJ_wash_pot}
\end{figure}

Tunneling weak links or Dayem nanobridges are examples of Josephson junctions (JJs). The former consist of two superconducting leads having a weak contact through a thin insulating layer, the latter are simply narrow short constrictions (bridges) in otherwise continuous superconducting material.
Supercurrent carrying state of a JJ or a superconducting nanowire is conveniently described within tilted washboard potential arising from the Resistively and Capacitavely Shounted Junction model (RCSJ)\cite{Martinis1988} (Fig.\,\ref{fig:JJ_wash_pot}). Within the model, state of the superconducting wavefunction is mapped into a position of a particle moving in the one-dimensional potential. The particle exhibits Brownian fluctuations due to interaction with constant temperature bath\cite{brownian_particle_Esteve}. It corresponds to random changes in the superconducting phase across the JJ around a mean value, meaning, by virtue of DC Josephson effect, average DC supercurrent flowing in the JJ. The height of the potential barrier separating two local minima is controlled by biasing current. For supercurrents much below critical current, the height of a potential barrier is much larger than accessible thermal energy \(k_BT\) and the particle can not escape through the barrier. However, increasing the biasing current, one can reduce the barrier height to an extent that thermal or quantum fluctuations are sufficient to drive the particle over the barrier\cite{Bezryadin2012,Zgirski2011,Dunkleberger1974}. If such a so-called phase slip happens\cite{Little1967,Langer1967}, the particle acquires sufficient inertia to jump over lower barriers (it is true for an underdamped junction). Superconducting wave function accumulates the phase and this, by virtue of AC Josephson effect, creates voltage across the JJ giving an experimentalist a mean to test the escape. We call such an event \textit{switching}. In case of superconducting wires and Dayem nanobridges, the voltage appears due to phase-slip followed by overheating and transition to normal state\cite{Tinkham2003, Pekola2008,Little1967}.
The Brownian behavior of superconducting wavefunction suggests that JJs and superconducting wires can be used as a random number generators. The probability for them to switch during rectangular current pulse of duration \textit{t} is \(P=1-\exp(-\Gamma t)\) with \(\Gamma\) being the escape rate dependent both on temperature and current amplitude (see Supplemental Material). This formula has general validity, both for wires and Josephson Junctions. For case of JJs the switching rate is \(\Gamma = (\omega_p/2 \pi )\exp(-\Delta U/k_B T_{esc})\), where \(\omega_p\) is natural frequency of the particle oscillations at the bottom of the potential, \(T_{esc}\) is an effective temperature of the escape\cite{Martinis1988}, \(\Delta U\) is the height of the potential barrier roughly equal to \(\Phi_0 I_c\) (\(\Phi_0\) - flux quantum, \(I_c\) - critical current) at zero current and can be lowered with bias current \(\Delta U(i)=\Delta U(0)(1-i/i_0)^{3/2}\). For 1D superconducting wires the formula is the same but \(\Delta U\) at zero current corresponds to the condensation energy of the smallest possible volume of the superconducting wire which can be driven normal i.e. \(\Omega = \xi S\)\cite{Langer1967} (\(\xi\) - superconducting coherence length, \textit{S} - wire cross-section) and exponent has the value of \(5/4\) instead of \(3/2\) \cite{Tinkham2003}. For thicker wires we can think of similar formula, but the exact energy landscape and switching scenario is more disputable (e.g. one could assume that energy fluctuation necessary to drive wire normal is related to the condensation energy of the piece of the wire of length \(\xi\)). Since it is not the primary goal of our paper we will not discuss it here but only notice that the exact nature of switching in thick superconducting wires is not essential for generating random bits.

\subsection{Protocol to generate random bits}
\begin{figure}[H]
  \centering
		\includegraphics[width=1.0\textwidth]{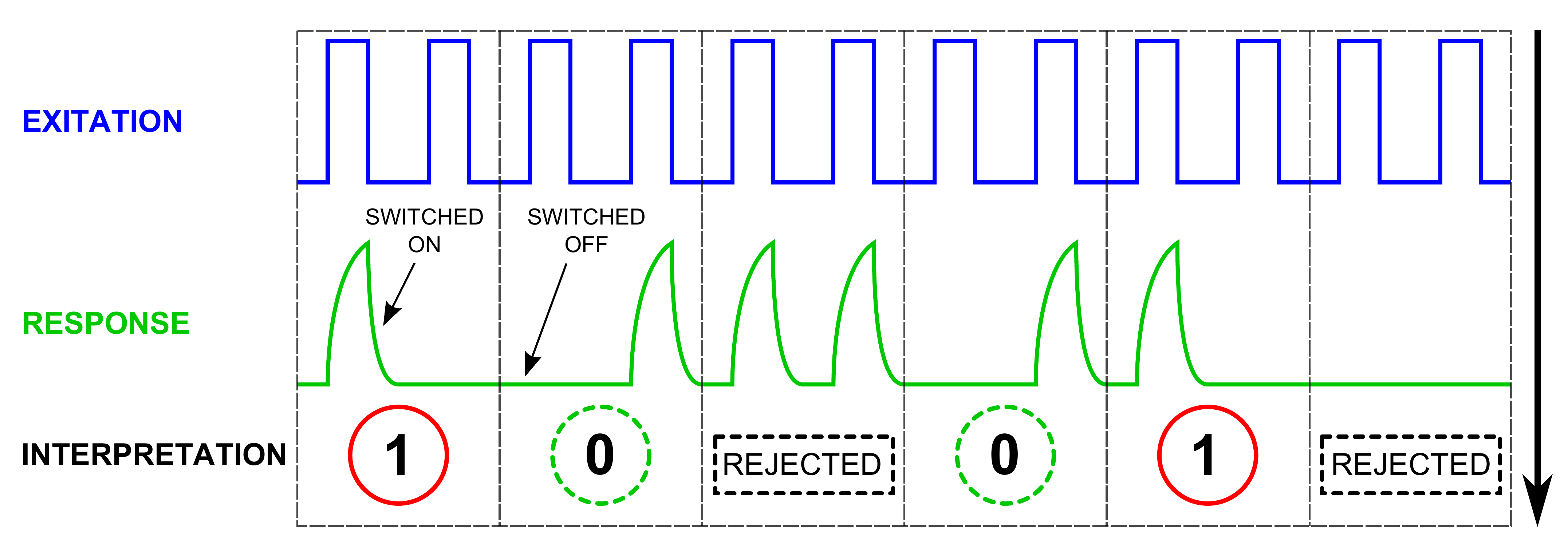}
  \caption{JJ tested with pulse train (EXCITATION). For each testing pulse the JJ switches or remains silent (RESPONSE). Response as recorded on oscilloscope is rounded, for it is measured with twisted pairs serving as low-pass filters. For analysis we split testing pulses in groups of 2. If within the group for first pulse the JJ switches and for second pulse does not, it encodes logical “1” (solid red circle). If it is opposite, logical “0” is encoded (dashed green circle). Different results (dotted black box) are discarded (INTERPRETATION).}
  \label{fig:process}
\end{figure}

Suppose we have a coin which is not fair in the sense that probability \textit{P} to get the head differs from 0.5. Moreover the probability varies very slowly with time. Obviously a game played with such a coin is not fair. However we can make it fair introducing following assumption: we flip the coin twice one flip after another. If in first flip we get the head and in the second the tail player A wins (logical “1”). If in first flip we get the tail and in the second the head player B wins (logical “0”). If two successive trials give the same result, the drawing is discarded. Since we assumed probability for flipping the head to vary slowly with time such a game can be considered fair, for it gives the same probability \(P(1-P)\) for both players to win after coin has been flipped twice. The procedure outlined above was first proposed by von Neuman\cite{Gifford1988} and is considered as a one of the most straightforward ways to unbias the random sequence i.e. to convert random sequence of zero and ones with unequal probabilities for both into random sequence for which probability to get bit 0 is the same as for bit 1, and equal 0.5.
JJ is a such not fair coin. It is difficult to set switching probability exactly equal to 0.5. But whatever this probability is, the response to 2 successive testing pulses encodes logical “0” or “1”, provided JJ switched for one testing pulse and did not switch for another. The idea of generating 4-bit random number is explained in Fig.\,\ref{fig:process}.
\subsection{Experimental}
\begin{figure}
  \centering	
		\includegraphics[width=0.9\textwidth]{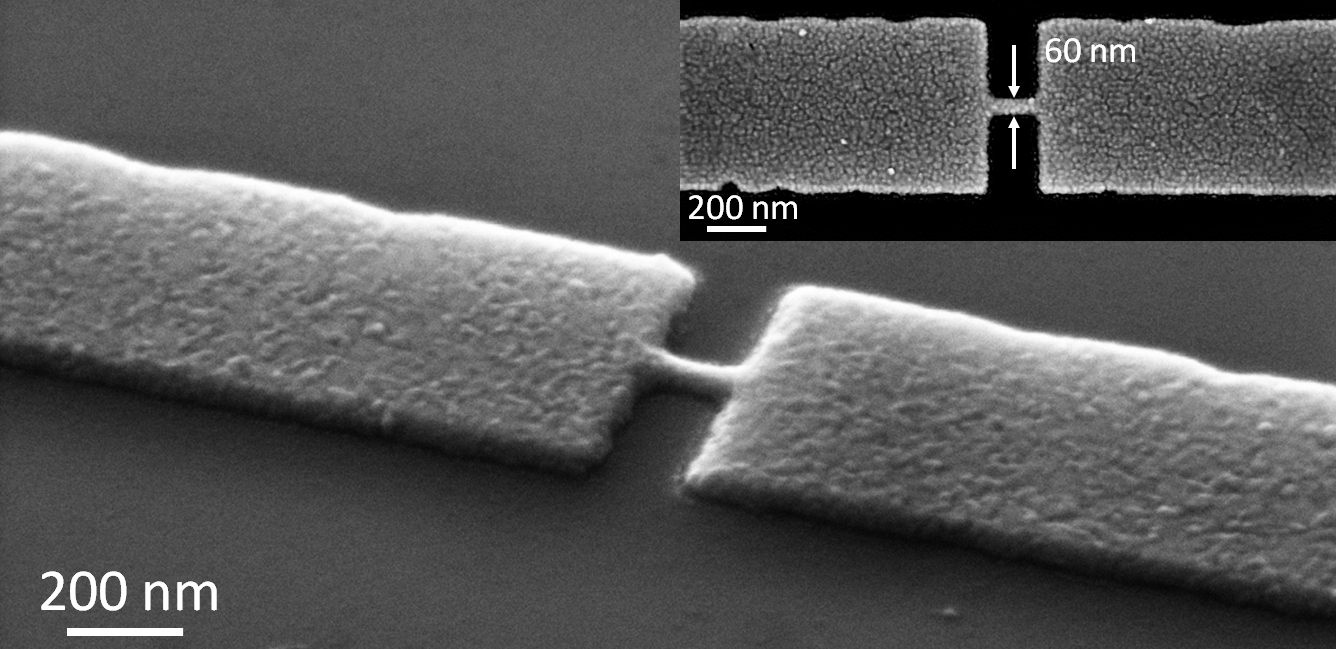}	
  \caption{Scanning electron micrograph of the Dayem nanobridge. Inset shows the nanobridge dimensions.}
  \label{fig:semphoto}
\end{figure}

We fabricated Dayem nanobridge by standard e-beam lithography followed by thermal evaporation of 30\,nm thick Aluminum (Fig.\,\ref{fig:semphoto}). The circuit employing the bridge as a random number generator is schematically indicated in Fig.\,\ref{fig:scheme}a. Details of the circuit are described in Methods (see Supplemental Material). The principle of operation is following. First, we record IV of the JJ by applying a triangular voltage sweep (Fig.\,\ref{fig:scheme}b). Secondly, we find current pulse amplitude for which JJ switches with probability of 0.5. It is easily accomplished by collecting a so-called S-curve (Fig.\,\ref{fig:scheme}c and Fig.\,\ref{fig:scheme}d).
The train of \(N_0\) current pulses is sent down the JJ and number of switchings \textit{n} is recorded. It gives switching probability \(P=n/N_0\) for a given current amplitude. Then current pulse amplitude is increased and the procedure is repeated. Having found the current amplitude \textit{A} (cf. Fig.\,\ref{fig:scheme}c) for which JJ switches with probability \(P\approx 0.5\) we have sent to JJ train of \(N_0=4\times 10^6\). We have recorded response of JJ with digitizer, collecting a point each 500ns. We can use such a low acquisition rate because the sustaining part of the pulse holds the memory of switching event over 5\(\mu s\). We have performed post-processing of the data in the spirit of idea explained in the Fig.\,\ref{fig:process}. On converting the data into sequence of zeros and ones we are ready to check its randomness.
\begin{figure}[H]
  \centering
		\includegraphics[width=0.5\textwidth]{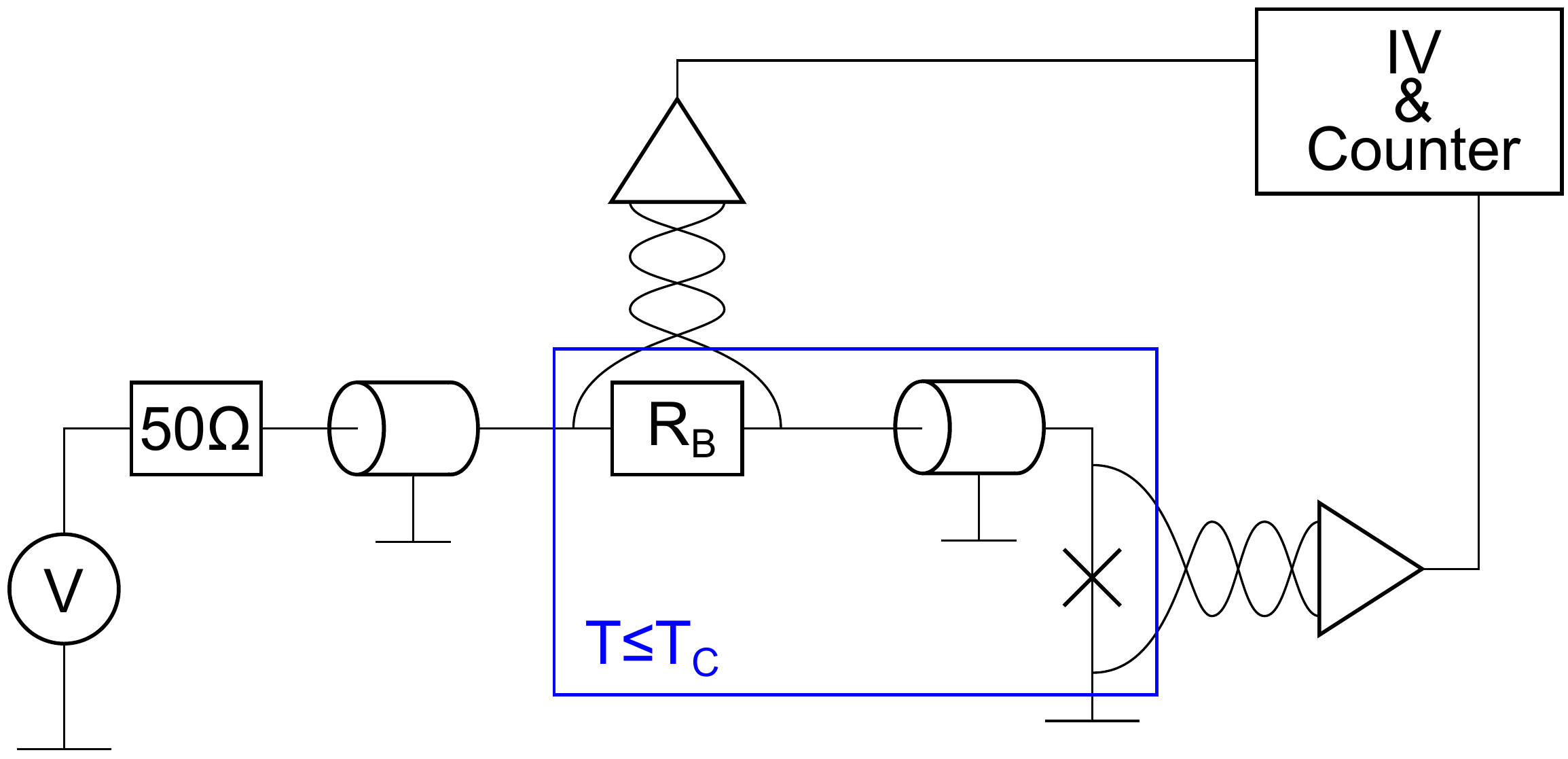}
			\put(-200,90){{(a)}}
		\includegraphics[width=0.45\textwidth]{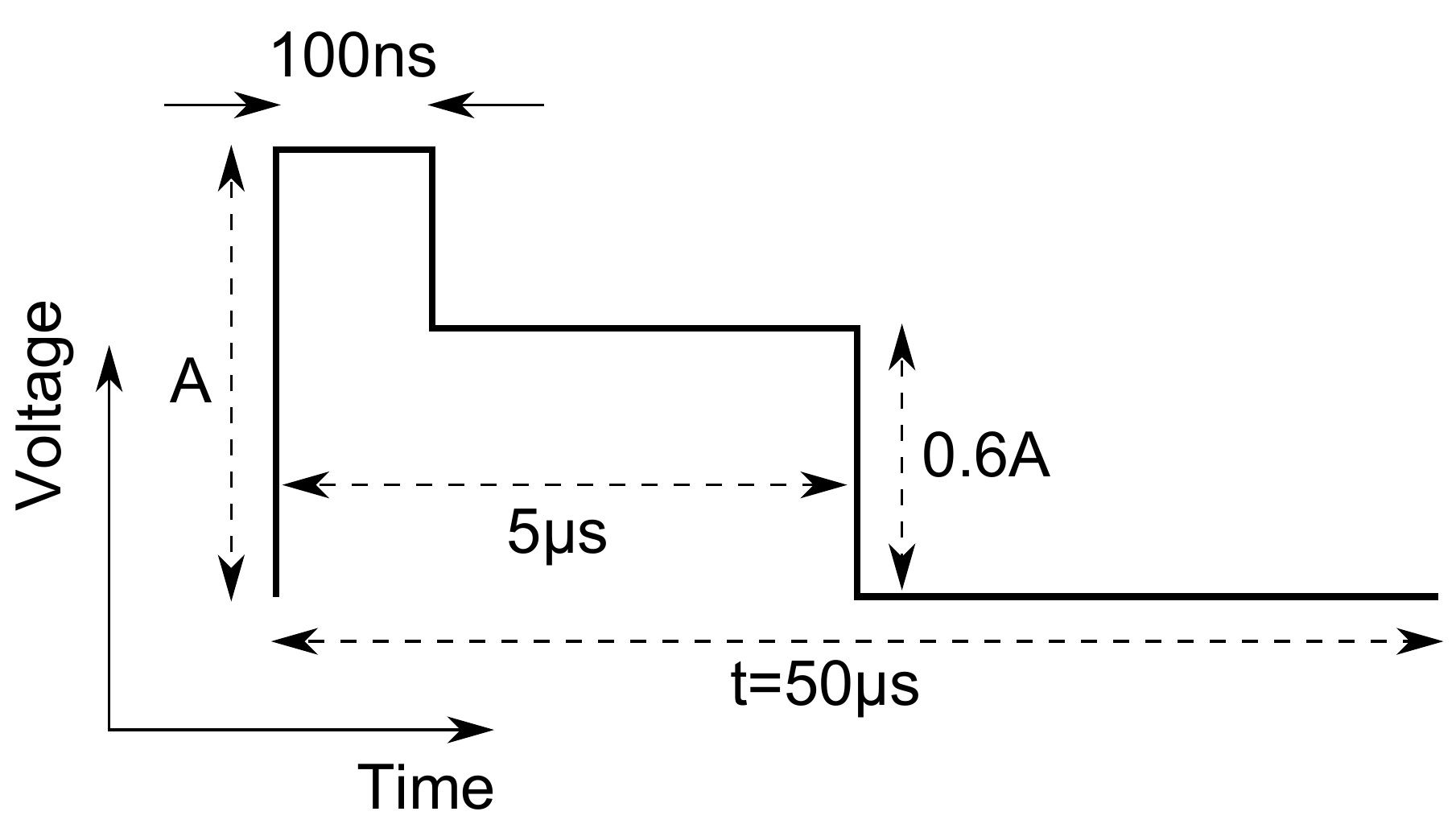}
			\put(-100,90){{(c)}}
\\*
		\includegraphics[width=1.0\textwidth]{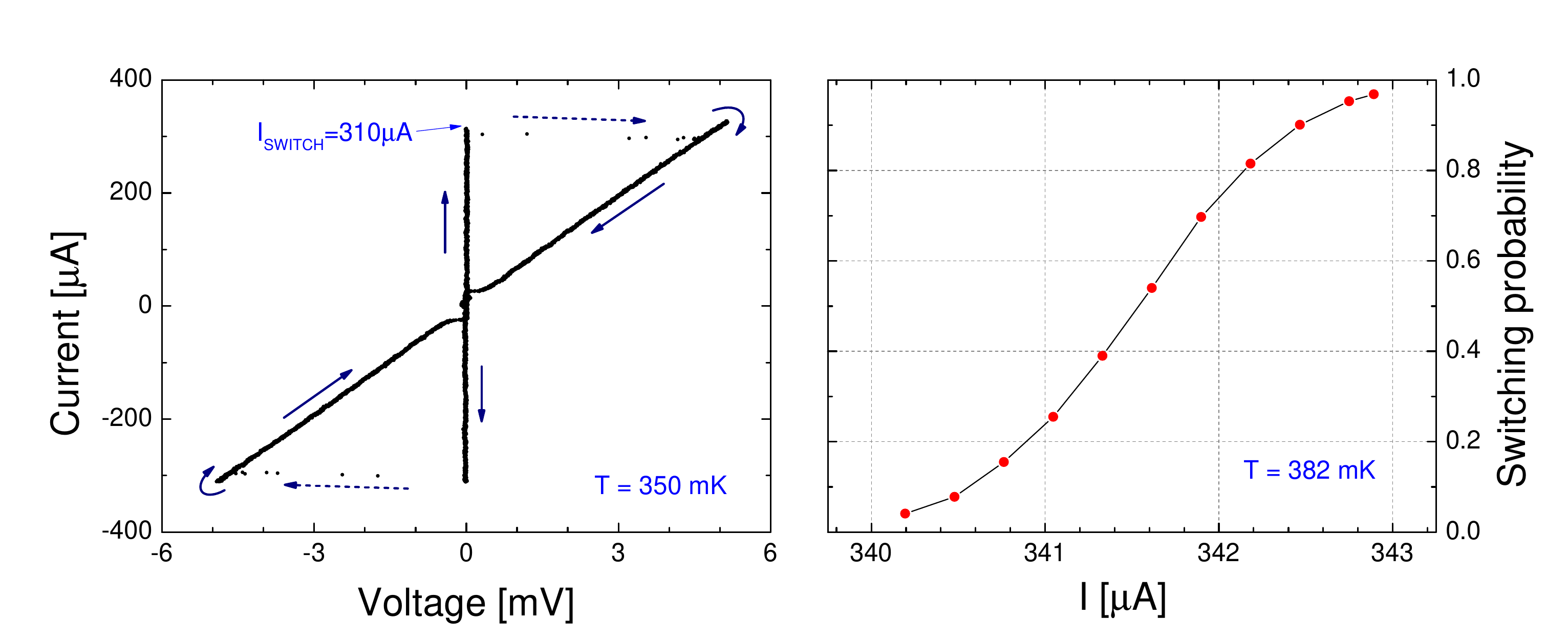}
			\put(-415,145){{(b)}}
			\put(-210,145){{(d)}}
  \caption{(a)\,The circuit employing a JJ for random number generation. (b)\,The Dayem nanobridge current-voltage characteristic revealing switching behavior at a threshold current. (c)\,Complex pulse used for JJ testing. Its amplitude \textit{A} defines probability for JJ to switch and lower plateau (sustaining part) allows for read-out with low-pass twisted pairs. (d)\,Experimentally obtained S-curve. Each point is the estimator for switching probability at given current amplitude A measured with train of \(N_0\)=10\,000 pulses. The line is a guide for the eye.}
  \label{fig:scheme}
\end{figure}
\subsection{Are generated bits random?}
We have generated a stream of \(N=10^6\) bits (see the \textit{Randombits.zip} attached in Supplemental Material). For non-biased sequence we expect to obtain \(0.5\cdot 10^6\) “ones” with standard deviation of \(\sqrt{NP(1-P)}=500\). We have obtained 500\,142\,“ones”. It allows us to proceed with more involved tests. There are many statistical tests for random sequences, but none quarantees 100\% certainty of randomness for lack of clear criteria and finite number of samples\cite{Drake1967}. Nevertheless we present a few statistical tests which seem to confirm the randomness of our stream.
\begin{figure}
  \centering
		\includegraphics[width=0.8\textwidth]{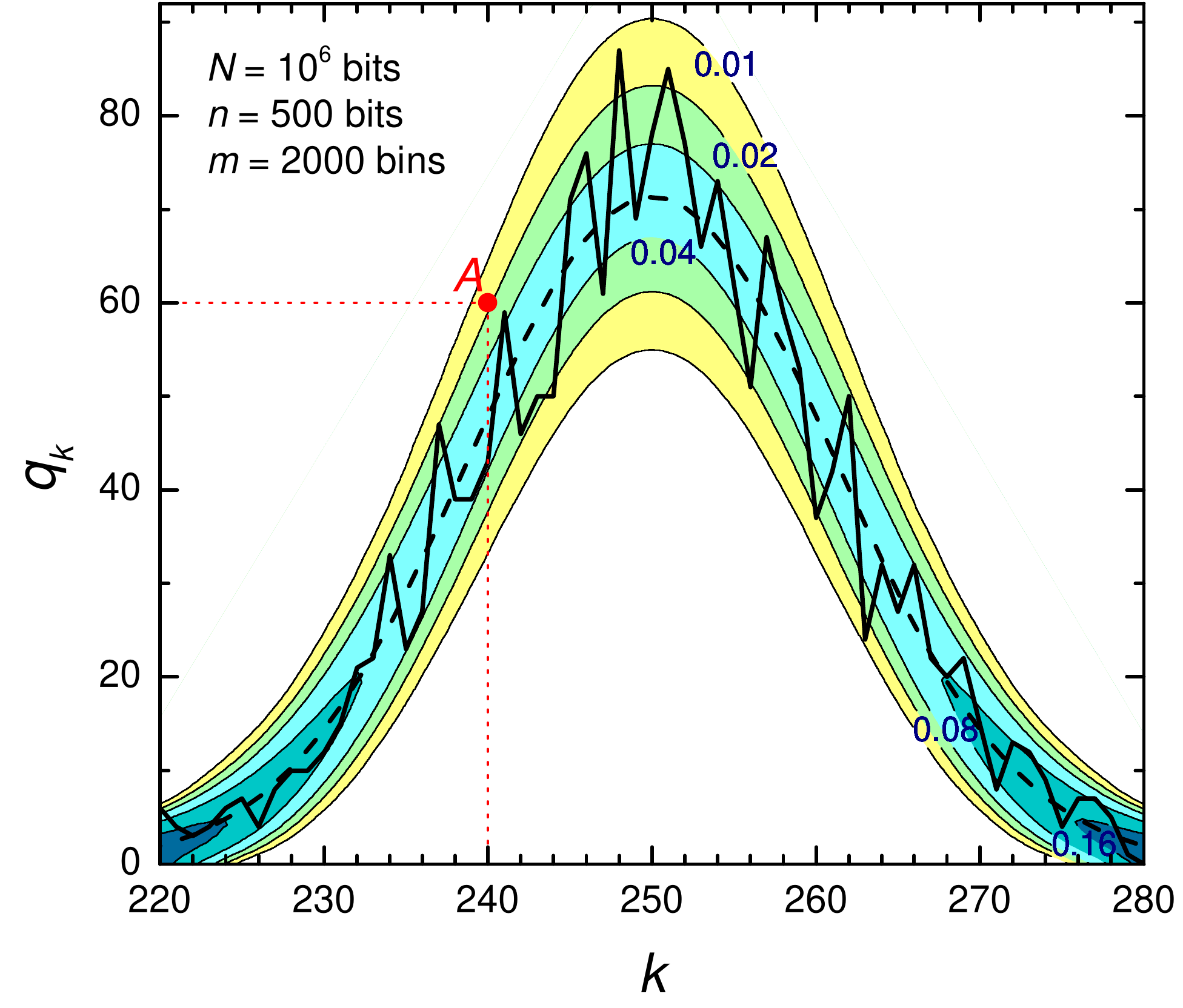}
  \caption{Probability map for getting \(q_k\) bins with \textit{k} “ones” with imposed experimentally determined distribution \(q^{\exp}_k\) – solid line. Dashed line represents \(<q_k>\) distribution. The interpretation of the probability map is explained for point A. It tells that probability for obtaining 60\,bins in total number of 2\,000\,bins, each with 240 “ones” is 0.02. We notice that experimentally determined distribution falls within the expected range of probabilities.}
  \label{fig:map}
\end{figure}

\setlength{\parindent}{1cm}First test starts with division of the sequence of \(N=n\cdot m\) samples (zeros and ones) into \textit{m} bins, each consisting of \textit{n} samples. Probability to obtain \textit{k} times “one” in a single bin of a length \textit{n} is given with binomial distribution: 
\begin{align}
	p_k={n \choose k}P^k (1-P)^{n-k} \stackrel{P=0.5}{=} {n \choose k}P^n
\end{align}
Its mean value is \(<k>=n\cdot P\) and standard deviation is \(\sigma = \sqrt{(1-P)P\cdot n}\).
Similarly probability to obtain \(q_k\) bins within m each with \textit{k} “ones” is:
\begin{align}
	p(q_k)={m \choose q_k}p^{q_k}_k (1-p_k)^{m-q_k}
\end{align}
The expected number of bins with \textit{k} “ones” is \(<q_k>=p_k\cdot m\). We expect the experimentally determined number of bins with \textit{k} “ones”, \(q^{\exp}_k\) to deviate on average from \(q_k\) by standard deviation \(\Delta q_k=\sqrt{p_k(1-p_k)m}\). In Fig.\,\ref{fig:map} we plot theoretical \(p(q_k)\) distribution subject to statistical broadening expected for the finite number of \textit{m} bins. In the same Figure experimentally determined distribution \(q^{\exp}_k\) is plotted. We conclude that the test does not negate the randomness of the sequence, for \(q^{\exp}_k\) has probability significantly different from zero for all \textit{k}–values and fluctuates around the mean value \(<q_k>\).

\begin{figure}
  \centering
		\includegraphics[width=1.0\textwidth]{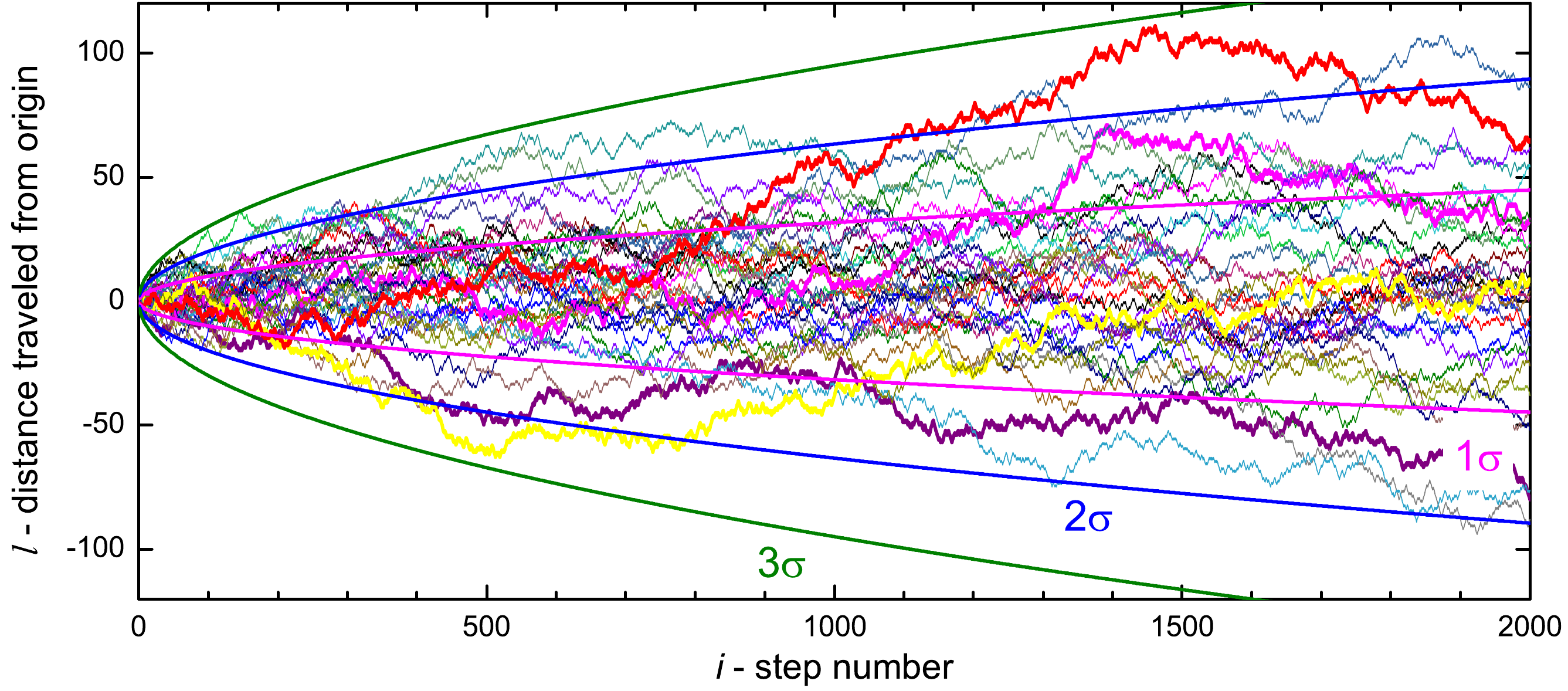}
		\put(-400,180){{\textbf{(a)}}}
\\*
		\includegraphics[width=1.0\textwidth]{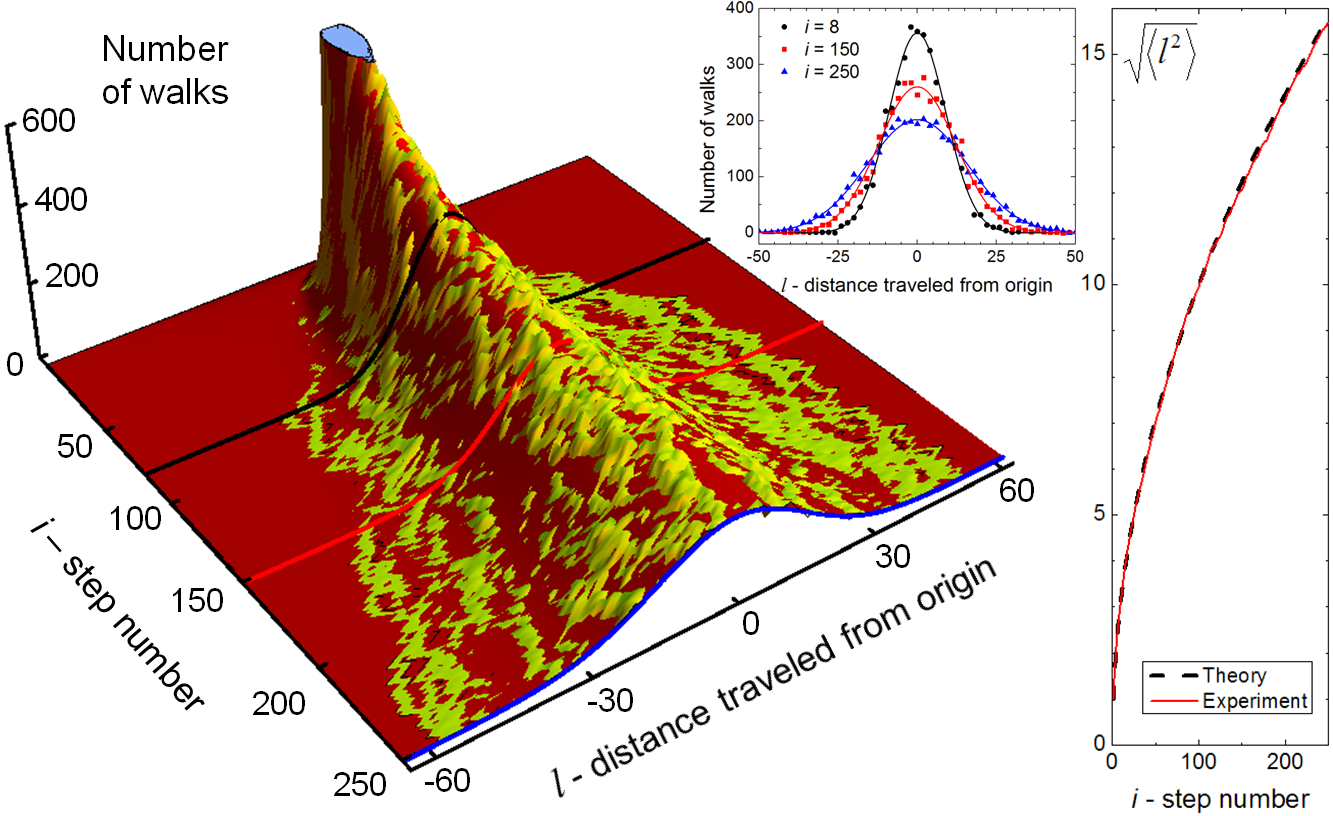}
		\put(-420,210){{\textbf{(b)}}}
		\put(-30,90){{\textbf{(c)}}}
  \caption{(a) Representative random walks with theoretical standard deviation imposed (\(\sigma, 2\sigma, 3\sigma\) curves). (b) Evolution of random walks distribution with step number \(i\). Smooth red surface is theory - Gaussian distributions with \(\sqrt{<l^2>}\) standard deviations. Yellow rough surface corresponds to experimental distribution established on summing of \textit{m}=4\,000 walks. Three sections, along solid lines, are chosen for clarity and presented in the separate plot. (c) Einstein-Smoluchowski law (dashed black line) compared with experimental data (solid red line).}
\label{fig:randomwalk}
\end{figure}

Second test utilizes the concept of random walk. We divide the stream of \textit{N} bits into \textit{m} bins. Each bin defines one random walk with bit “1” meaning one step forward and bit “0” meaning one step backward. Such a walk, if really random, should obey Einstein-Smoluchowski law: \(<l^2>=i\), where \textit{l} is a distance traveled from the origin after \textit{i} steps. The movement corresponds to 1D diffusion of a particle. The distance traveled by the particle after \textit{i} steps is described with Gaussian distribution with mean value 0 and variance \(<l^2>\). We present trajectories of numerous walks in Fig.\,\ref{fig:randomwalk}a. On imposing all walks on each other (Fig.\,\ref{fig:randomwalk}b) we obtain a distribution of final positions of the particle after \(1\), \(2\), ..., \(i\), ..., \(n\) steps. In Fig.\,\ref{fig:randomwalk}c we compare average deflection for walks \(<l^2>\) after \textit{i} steps against Einstein-Smoluchowski law. We conclude that the walks are indeed random.

In the third test we have calculated autocorrelation of our bits and found no obvious evidence of frequency components (see Supplemental Material). Randomness of our data is also confirmed by NIST Test Suite\cite{NIST} (see Supplemental Material).

\subsection{Discussion and possible improvements}
We have conducted analogous experiment on superconducting Aluminum nanowire (30\,nm thickness, 600\,nm cross-section). Sequences obtained for the nanowire also pass tests for randomness. However, since switching current for our nanowire is higher than for Dayem nanobridge, it takes longer time for the nanowire to recover after switching to dissipative state. Switching produces a number of quasiparticles\cite{Zgirski2011}. It accounts for rising the temperature and increases the switching probability in the next trial. However by using a tunnel JJ we can switch to a finite voltage with a very small current (due to multiple Andreev reflections\cite{Urbina1997}), and consequently a small power dissipated in the JJ. Another approach is to use prepulse, preceding actual measuring pulse. Due to larger amplitude (say 1.3 of that of the measuring pulse) it makes the JJ switch (so called forced switching) and nulls a memory of the JJ. One can say that on average after forced switching the JJ is left with the same number of quasiparticles. The forced switching removes a possible correlation between two successive trials – the obligatory requirement for a good random number generator.

One can envisage generation of random bits with magnetic clusters. Magnetization reversal in ferromagnetic nanoclusters, if thermally excited, is described with the Neel-Brown model\cite{Wernsdorfer1997}. The picture of magnetization reversal in the model remains in the complete analogy to the JJ escape out of the metastable state. It follows one can test stochastic character of magnetization reversal by the same measuring protocol we have presented, but rather than pulses of current, magnetic field pulses should be used to give the magnetization a chance to reverse.

\subsection{Conclusions}
In conclusion, we have presented an original way to generate the random bit series exploiting inherent randomness of the switching from a superconducting to a non-zero voltage state in Josephson junctions and superconducting nanowires. In our not-yet-optimised nanodevices we have achieved random number generation rates of 10-100\,kb/s. Our experiments have shown that Cooper pairs in these systems exhibit collective response to a random external stimulus, which allows to treat them as a single archetypal Brownian particle.

\begin{acknowledgement}
The authors thank Tomasz Dietl for his helpful advice, Łukasz Cywiński for discussions, and Cezary Śliwa for a technical support. We are very grateful to the Foundation for Polish Science for funding this work through the HOMING PLUS program. We also thank National Science Center, grant MAESTRO (2011/02/A/ST3/00125), and the EAgLE Project. We acknowledge the National Institute of Standards and Technology for access to Test Suite Software.
\end{acknowledgement}


\bibliography{Gambling_with_superconducting_fluctuations}

\end{document}